\documentclass[sigconf]{acmart}

\usepackage{booktabs} 

\usepackage{hyperref}

\definecolor{linkcolor}{RGB}{19,83,144}

\hypersetup{
    colorlinks=true,
    linkcolor=linkcolor,
    filecolor=linkcolor,      
    urlcolor=linkcolor,
    citecolor=linkcolor
}

\newcommand{\cwelink}[1]{\href{https://cwe.mitre.org/data/definitions/#1.html}{CWE-#1}}

\copyrightyear{2020} 
\acmYear{2020} 
\setcopyright{licensedusgovmixed}\acmConference[ACSAC 2020]{Annual Computer Security Applications Conference}{December 7--11, 2020}{Austin, USA}
\acmBooktitle{Annual Computer Security Applications Conference (ACSAC 2020), December 7--11, 2020, Austin, USA}
\acmPrice{15.00}
\acmDOI{10.1145/3427228.3427257}
\acmISBN{978-1-4503-8858-0/20/12}

\begin{document}

\title[]{Measurements of the Most Significant\\ Software Security Weaknesses}


\author{Carlos Cardoso Galhardo}
\affiliation{%
  \institution{National Institute of Standards and Technology; INMETRO}
}
\email{carlos.cardosogalhardo@nist.gov; cegalhardo@inmetro.gov.br}

\author{Peter Mell}
\orcid{0000-0003-2938-897X}
\affiliation{%
  \institution{National Institute of Standards and Technology}
}
\email{peter.mell@nist.gov}

\author{Irena Bojanova}
\affiliation{%
  \institution{National Institute of Standards and Technology}
}
\email{irena.bojanova@nist.gov}

\author{Assane Gueye}
\affiliation{%
  \institution{UADB-Senegal \& Prometheus Computing}
}
\email{assane1.gueye@uadb.edu.sn}

\renewcommand{\shortauthors}{C. Galhardo et al.}

\begin{abstract}
In this work, we provide a metric to calculate the most significant software security weaknesses as defined by an aggregate metric of the frequency, exploitability, and impact of related vulnerabilities. The Common Weakness Enumeration (CWE) is a well-known and used list of software security weaknesses. The CWE community publishes such an aggregate metric to calculate the `Most Dangerous Software Errors'. However, we find that the published equation highly biases frequency and almost ignores exploitability and impact in generating top lists of varying sizes. This is due to the differences in the distributions of the component metric values. To mitigate this, we linearize the frequency distribution using a double log function. We then propose a variety of other improvements, provide top lists of the most significant CWEs for 2019, provide an analysis of the identified software security weaknesses, and compare them against previously published top lists.
\end{abstract}

%
%



\maketitle

\section{Introduction}
\label{sec.Intro}

In 2019, there were over 17\,000 documented software vulnerabilities \cite{NVD} that enable malicious activity. While many are discovered, they map to a relatively small set of underlying weakness types. We posit that if the most significant of these types can be identified, developers of programming languages, software, and security tools can focus on preventing them and thus over time diminish the quantity and severity of newly discovered vulnerabilities.

In this work, we provide a metric to calculate the most significant security weaknesses (MSSW) in software systems. We define a `significant' weakness as one that is both frequently occurring among the set of publicly published vulnerabilities and results in high severity vulnerabilities (those that are easily exploitable and have high impact). 
The set of security weakness types upon which we calculate significance comes from
the Common Weakness Enumeration (CWE) \cite{CWE}.
We also leverage the Common Vulnerabilities and Exposures (CVE) \cite{CVE} repository of publicly announced vulnerabilities, the Common Vulnerability Scoring System (CVSS) \cite{CVSSv3.1} to measure the severity of vulnerabilities, and the National Vulnerability Database (NVD) \cite{NVD} to map the CVEs to both CWEs and CVSS scores.

In the fall of 2019, the CWE community published an equation to calculate the `Top 25 Most Dangerous Software Errors' (MDSE) among the set of CWEs \cite{MitreTop25}. It follows the form of the common security risk matrix combining probability and severity (e.g., \cite{engert1999risk}). The MDSE equation claims to combine `the frequency that a CWE is the root cause of a vulnerability with the projected severity'; the equation description implies that both factors are weighed equally (making no mention of any bias). However, we empirically find that the equation highly biases frequency and almost ignores severity in generating top lists of varying sizes. This is due to the equation multiplying calculated frequency and severity values together though each has has very different distributions. Frequency distributions have a power law like curve, while severity distributions are more uniform. Our mitigation is to create a revised equation, named MSSW, that adjusts the frequency distribution using a double log function to better match it to the severity distribution. We also fix an error in how normalization is done in the MDSE equation.

We next improve upon the data collection approach used by the MDSE equation by leveraging published literature \cite{mell2020}. Lastly, we publish top lists of the most significant CWEs for 2019, provide an analysis of those software security weaknesses, and compare our top lists against previously published lists. It is our hope that our data and methodology will be adopted to focus our collective security resources in reducing the most significant software security weaknesses.

The rest of this work is organized as follows. Section \ref{sec.Background} provides background on CVE, CVSS, CWE, NVD, and the MDSE equation. Section \ref{sec.Limitations} discusses the limitations of the MDSE equation. Section \ref{sec.FixedEquation} presents our MSSW equation that mitigates the previously identified limitations. Section \ref{sec.TopWeaknesses} provides two lists of the most significant CWEs at two different levels of software flaw type abstractions. Section \ref{sec.Evaluation} provides a discussion and analysis of the most significant CWEs identified. Section \ref{sec.RelatedWork} presents related work, Section \ref{sec.FutureWork} discussed possible future research, and Section \ref{sec.Conclusions} concludes. 


\section{Background}
\label{sec.Background}

\subsection{Common Vulnerabilities and Exposures}
The CVEs are a large set of publicly disclosed vulnerabilities in widely-used software. They are enumerated with a unique identifier, described, and referenced with external advisories \cite{CVE} \cite{baker1999CVE}. 

\subsection{Common Vulnerability Scoring System}
CVSS `provides a way to capture the principal characteristics of a vulnerability and produce a numerical score reflecting its severity' \cite{CVSS}. The CVSS base score reflects the inherent risk of a vulnerability apart from any specific environment. 
The base score is composed from two sub-scores that calculate exploitability (how easy it is to use the vulnerability in an attack) and impact (how much damage the vulnerability can cause to an affected component). 

The exploitability score is determined by the following:
\begin{itemize}
    \item attack vector: `the context by which vulnerability exploitation is possible',
    \item attack complexity: `the conditions beyond the attacker’s control that must exist in order to exploit the vulnerability',
    \item privileges required: `the level of privileges an attacker must possess before successfully exploiting the vulnerability', and
    \item user interaction: a human victim must participate for the vulnerability to be exploited.
\end{itemize}

The impact score is determined by measuring the impact to the confidentiality, integrity, and availability of the affected system. Also included is a scope metric that `captures whether a vulnerability in one vulnerable component impacts resources in components beyond its security scope'. The specifics on these metrics and the details for the three equations can be found in the CVSS version 3.1 specification at \cite{CVSSv3.1}.

\subsection{Common Weakness Enumeration}
\label{CWE}

The Common Weakness Enumeration (CWE) \cite{martin2008CWE} is a `community-developed list of common software security weaknesses'. `It serves as a common language, a measuring stick for software security tools, and as a baseline for weakness identification, mitigation, and prevention efforts' \cite{CWE}. It contains an enumeration, descriptions, and references for 839 software weaknesses that are referred to as CWEs, where each is labelled CWE-\textit{X} with \textit{X} being an integer. 

The CWE weaknesses model has four layers of abstraction: pillar, class, base, and variant. There is also the notion of a compound, that associates two or more interacting or co-occurring CWEs \cite{MitreGlossary}. These abstractions reflect to what extent issues are described in terms of five dimensions: behavior, property, technology, language, and resource. Variant weaknesses are at the most specific level of abstraction; they describe at least three dimensions. Base weaknesses are more abstract than variants and more specific than classes; they describe two to three dimensions. Class weaknesses are very abstract; they describe one to two dimensions, typically not specific about any language or technology. Pillar weaknesses are the highest level of abstraction.


There are a set of taxonomies, called views, to help organize the CWEs. Two prominent CWE taxonomies are the `Research Concepts' (view 1000) and `Development Concepts' (view 699). There is also a view 1003 that was made specifically to describe the set of CVEs that contains 124 CWEs. It is called `CWE Weaknesses for Simplified Mapping of Published Vulnerabilities View'.

\subsection{National Vulnerability Database}
The CWE effort uses the National Vulnerability Database (NVD) \cite{NVD} as a repository of data from which to calculate the MDSE scores. The NVD contains all CVEs and for each CVE it provides a CVSS score along with the applicable CWE(s) that describe the weakness(es) enabling the vulnerability. For the empirical work in this paper, we use the complete set of
17\,308
CVEs published by NVD for 2019, that were available as of
2020-03-19. 

\subsection{Most Dangerous Software Error Equation}
The MDSE equation is designed to balance the frequency and severity in ranking the CWEs. The frequency is determined by the number of CVEs that map to a given CWE in the time period of study. The severity is determined by the mean CVSS score for the CVEs mapped to a given CWE. The MDSE score for a CWE is produced by multiplying the normalized frequency by the normalized severity and then multiplying by 100. We now describe this metric more formally.

\subsubsection{Metric for Normalized Frequency}
\label{MDSE-Frequency}

Let $I$ designate the set of all CWEs and let $J$ be the set of all CVEs.

For CWE $i\in I$, let $N_i$ be the number of CVEs mapped to $i$, defined as follows:

\begin{equation}
N_i = \sum_{j\in J} e_{ij},
\label{eq:N_i}
\end{equation}
where 

\begin{equation}
e_{ij} =
    \begin{cases}
          1, &         \text{if CVE $j$ is mapped to CWE $i$},\\
            0, &         \text{otherwise}.
    \end{cases}
    \label{eq:eij}
\end{equation}

Now let $F_i$ be the normalized frequency for CWE $i$, defined as follows:

\begin{equation}
F_i = \frac{N_i - \min\limits_{i' \in I} (N_{i'})}{\max \limits_{i' \in I} (N_{i'}) - \min\limits_{i' \in I} (N_{i'})}.
\label{eq:F_i}
\end{equation}


\subsubsection{Metric for Normalized Severity}
\label{MDSE-Severity}

Let $J$, $N_i$, and $e_{ij}$ be as defined above in Section \ref{MDSE-Frequency}.
Let $s_j$ be the CVSS base score for CVE $j$. For CWE $i\in I$, let $\overline{S_i}$ be the mean CVSS score, defined as follows:

\begin{equation}
\overline{S_i} =
\frac{\sum_{j\in J}s_j e_{ij}}{N_i}.
\label{eq:S_i_overline}
\end{equation}

Now let $S_i$ be the normalized severity for CWE $i$, defined as follows:

\begin{equation}
S_i = \frac{\overline{S_i} - \min\limits_{j \in J} (s_j)}{\max \limits_{j \in J} (s_j) - \min\limits_{j \in J} (s_j)}.
\label{eq:S_i}
\end{equation}

\subsubsection{Most Dangerous Software Error Metric}
Let $MDSE_i$ be the MDSE score for CWE $i$, defined as follows:

\begin{equation}
MDSE_i=F_i*S_i*100.
\label{eq:MDSE_i}
\end{equation}



\section{Limitations of the Equation}
\label{sec.Limitations}

\begin{figure}
\centerline{\includegraphics[scale=.6]{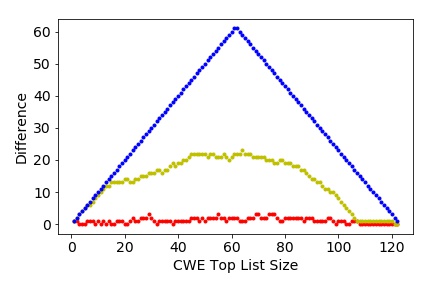}}
\caption{The Size of the Set Difference between Top Lists from the MDSE Equation Compared to Frequency Top Lists (red bottom line), Severity Top Lists (yellow middle line), and the Theoretical Maximum (blue top line)}
\label{fig:MDSE-Set-Difference}
\end{figure}

The MDSE equation was designed to and appears to combine both frequency and severity in determining the individual scores used to rank the CWEs. The frequency component is calculated in equation \ref{eq:F_i} and the severity component is calculated in equation \ref{eq:S_i}; both are brought together in equal proportions in equation \ref{eq:MDSE_i} to create the MDSE score. And both the severity and frequency are normalized in equations \ref{eq:F_i} and \ref{eq:S_i} to ensure that their scales match for the multiplication in equation \ref{eq:MDSE_i}.

However, we empirically find that the MDSE equation strongly biases frequency over severity. To demonstrate this, we calculate MDSE top CWE lists for all possible list sizes. While there exist 839 CWEs, the CVE data used as MDSE input is mapped only to 124 view 1003 CWEs (see section \ref{CWE})\footnote{This is expected as view 1003 was designed to cover the types of vulnerabilities in CVE.
}. Thus the maximum top list size is 124. We also calculate top CWE lists using just the frequency equation \ref{eq:F_i} and then just the severity equation \ref{eq:S_i}. For each CWE top list size, we perform a set difference between the MDSE top list and the frequency top list. We then also do this between the MDSE top list and the severity top list. The size of the set difference between the MDSE top list and the frequency top list (for all possible top list sizes) has a maximum difference of 3. The size of the set difference between the MDSE top list and the severity top list (for all possible top list sizes) has a maximum difference of 23. This is shown graphically in Figure \ref{fig:MDSE-Set-Difference}. The bottom red line represents the set difference using frequency and the yellow middle line represents the set difference using severity. The top blue line shows the maximal possible set difference that could be achieved using the 124 CWEs. 

More qualitatively, the red line hovers close to a y-axis value of 0 which means that for all list sizes the top list generated using just frequency is almost identical to the top list generated using the MDSE equation. The middle yellow line being far from the y-axis value of 0 means that for all list sizes the top list generated using just severity is very different from the top list generated using the MDSE equation. Note that the yellow line shows an almost maximal difference for top list sizes of up to 15. 

\begin{figure}
\centerline{\includegraphics[scale=.6]{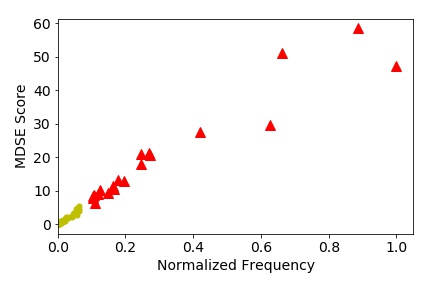}}
\caption{CWEs Chosen (Red Triangles) and Not Chosen (Yellow Circles) for a MDSE Top 20 List Relative to Frequency}
\label{fig:MDSE-vs-Frequency}
\end{figure}

\subsection{Limitation 1: Distribution Differences}
\label{lim1}
The MDSE equation in practice biases frequency over severity, even though its equations treat them equally, because frequency and severity have very different distributions. The frequency distribution has the majority of CWEs at a very low frequency and a few at a very high frequency (somewhat resembling a power law curve). This can be seen in Figure \ref{fig:MDSE-vs-Frequency} by looking at how each CWE maps to the x-axis (note that most of the yellow dots overlap, there are 102 yellow dots and 20 red triangles). The figure shows the MDSE scores for each CWE and shows how (for a top list of size 20) the top scoring chosen CWEs are exactly the most frequent CWEs. This is not unique and occurs for many top lists (e.g., for sizes 11, 13, 15, 16, 20, 21, 32, and 38) as shown when the bottom red line is at 0 in Figure \ref{fig:MDSE-Set-Difference}. The other sizes of top lists produce graphs that are almost identical to that in Figure \ref{fig:MDSE-vs-Frequency}, with at most 3 yellow circles just to the right of the leftmost red triangles representing the chosen CWEs. 
\begin{figure}
\centerline{\includegraphics[scale=.6]{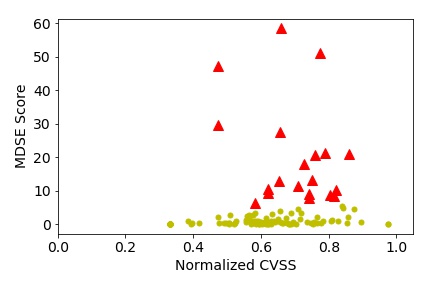}}
\caption{CWEs Chosen (Red Triangles) and Not Chosen (Yellow Circles) for a MDSE Top 20 List Relative to Severity}
\label{fig:MDSE-vs-CVSS}
\end{figure}

The severity distribution is more uniform within a limited range. It can be seen in Figure \ref{fig:MDSE-vs-CVSS} by looking at how the CWEs map to the x-axis. This figure shows how the top MDSE scoring chosen CWEs do not necessarily map to the CWEs with the highest severity. In fact, only 1 of the top 10 most severe CWEs made the MDSE top 20 list (note that many of the yellow circles lay on top of each other).

\subsection{Limitation 2: Normalization Error}
\label{lim2}
Equation \ref{eq:S_i} normalizes $S_i$ based on the maximum and minimum CVSS score found in the set of inputted CVEs. However, this does not lead to the expected and desired normalized distribution from 0 to 1. For our data the range is from .28 to .97, as can be seen from the mappings of the points onto the x-axis in Figure \ref{fig:MDSE-vs-CVSS}. The reason for this is that  $\overline{S_i}$ has a smaller range than the maximum and minimum CVSS score because each $\overline{S_i}$ represents the mean of the CVSS score for the CVEs that map to CWE $i$. This limitation, while of less consequence than the previous, constrains the range of $S_i$ values thus further lessening the influence that severity has in determining a MDSE score.


\section{Mitigated Equation}
\label{sec.FixedEquation}

We mitigate the limitations of the MDSE equation by replacing equations \ref{eq:F_i}, \ref{eq:S_i}, and \ref{eq:MDSE_i} with the five equations that follow:

\begin{equation}
k = \frac{1}{\log _{e} \log _{e} \max\limits_{i \in I} (N_{i}) },
\label{eq:k}
\end{equation}

\begin{equation}
F'_i =
    \begin{cases}
          \log _{e} N_i, & \text{if $N_i$ >= 1},\\
          0, & \text{otherwise},
    \end{cases}
\label{eq:F'_i}
\end{equation}

\begin{equation}
F''_i =
    \begin{cases}
          k\log _{e} F'_i, & \text{if $F'_i$ >= 1},\\
          0, & \text{otherwise},
    \end{cases}
\label{eq:F''_i}
\end{equation}

\begin{equation}
S'_i = \frac{\overline{S_i} - \min\limits_{i' \in I} (\overline{S_{i'}})}{\max \limits_{i' \in I} (\overline{S_{i'}}) - \min\limits_{i' \in I} (\overline{S_{i'}})},
\label{eq:S'_i}
\end{equation}

\begin{equation}
MSSW_i=F''_i*S'_i*100.
\label{eq:MSSW_i}
\end{equation}


\subsection{Explanation of Mitigated Equation}

\begin{figure}
\centerline{\includegraphics[scale=.6]{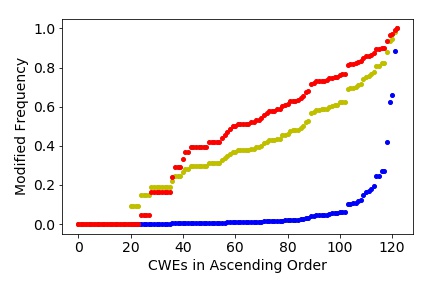}}
\caption{Normalized Distributions of Frequency (bottom blue line), Log of Frequency (middle yellow line), and Double Log of Frequency (top red line).}
\label{fig:G0-distributions}
\end{figure}

Equation \ref{eq:F'_i} takes the log of the frequency using the natural log as the base. Equation \ref{eq:F''_i} then takes the log of equation \ref{eq:F'_i}, again using the natural log as a base and multiplies the result by $k$ (from equation \ref{eq:k}). The $k$ coefficient serves the purpose of normalizing the resulting values between 0 and 1 (to match the severity range in equation \ref{eq:S'_i}).

These three equations modify the power law like frequency distribution to make it more linear, thus addressing limitation 1 (from Section \ref{lim1}). This can be seen in Figure \ref{fig:G0-distributions}. Each value on the x-axis represents a particular CWE, ordered from least frequent to most frequent. The lower blue line represents the normalized frequency (i.e., number of CVEs mapped to a particular CWE). Note the slow increase in frequency up to the 100th CWE, followed by a rapid increase terminating in an almost vertical line (i.e. large derivative). This behavior creates large differences between the most frequent CWEs and almost no difference between the lowest CWEs.  If $g(x) = \log f(x)$ its derivative is  
\begin{equation}
\frac{dg(x)}{dx} = \frac{1}{f(x)} \frac{df(x)}{dx},    
\end{equation}
thus applying a log function over the frequency should minimize differences between the most frequent CWEs.

The middle yellow line represents taking the log of the frequency (equations not shown), which helps linearize but still results in an upwards curve on the right side. Thus, we apply a double log for further linearization (see the top red line). We note that this approach is not pseudo-linear for the most infrequent of CWEs. However, this does not cause problems as our goal is to identify the most significant and any such CWE must have at least a moderate frequency.


Our modified MDSE equation \ref{eq:MSSW_i} then multiplies frequency and severity as in the original MDSE equation, but it multiplies from two distributions that have a similar shape for the part of the functions that are of interest. This enables the MSSW equation to more fairly balance evaluating frequency and severity in scoring and ranking a CWE. 

To address limitation 2 from Section \ref{lim2}, equation \ref{eq:S'_i} normalizes the severity using the maximum and minimum mean severity values. This gives the distribution a full 0 to 1 range which is not achieved in the MDSE equation \ref{eq:S_i}.

Equation \ref{eq:MSSW_i} is our final modified MDSE equation. We recommend its use in place of the published MDSE equation.


\subsection{Analysis of Mitigated Equation}

We now conduct three experiments to evaluate the effect of the MSSW equation in making the frequency and severity distributions more similar and in producing top lists with more equal inclusion of both frequency and severity. A fourth experiment involving correlation calculations is provided in Section \ref{sec.TopWeaknesses} (because it includes some variants introduced in that section).

\subsubsection{Risk Map Experiment} 

\begin{figure}
\centerline{\includegraphics[scale=.6]{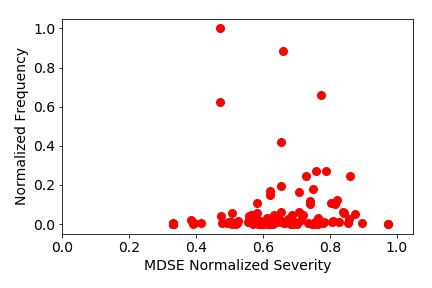}}
\caption{MDSE Equation Risk Map}
\label{fig:MDSE-Riskmap}
\end{figure}

\begin{figure}
\centerline{\includegraphics[scale=.6]{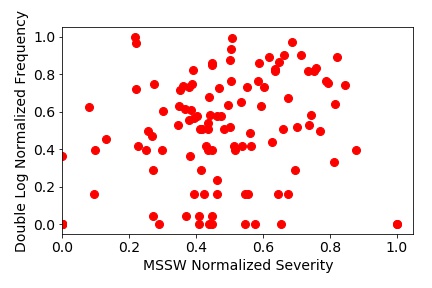}}
\caption{MSSW Equation Risk Map}
\label{fig:MSSW-Riskmap}
\end{figure}

Figure \ref{fig:MDSE-Riskmap} shows an MDSE risk map for the evaluated CWEs. Each red dot represents a CWE positioned according to its $S_i$ severity and $F_i$ frequency. In general, CWEs towards the upper right are more significant and those towards the lower left are less significant. Note how the majority of the CWEs are squished very close to the x-axis as many have a very small frequency. Also, the range of x-values is constrained from .37 to .97 (when the normalization should make it from 0 to 1).

Figure \ref{fig:MSSW-Riskmap} shows the same risk map using our double log frequency $F_i''$ and our modified severity $S_i'$. Note how the CWEs are now more uniformly spread over the y-axis. Also, the range of x-axis values is now from 0 to 1. The MSSW equation that combines frequency and severity using the values shown in Figure \ref{fig:MSSW-Riskmap} will now more equally combine them than with the MDSE values shown in Figure \ref{fig:MDSE-Riskmap}. 

\subsubsection{Set Difference Experiment} 

\begin{figure}
\centerline{\includegraphics[scale=.6]{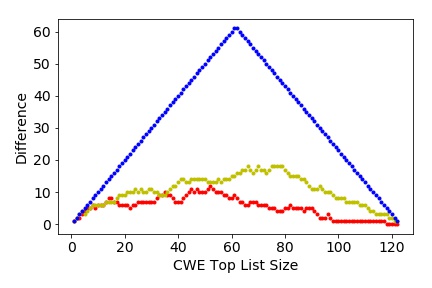}}
\caption{The Size of the Set Difference between Top Lists from the MSSW Equation Compared to Frequency Top Lists (red lower line), Severity Top Lists (yellow middle line), and the Theoretical Maximum (blue top line)}
\label{fig:MSSW-Set-Difference}
\end{figure}

In Figure \ref{fig:MSSW-Set-Difference} we show the size of the set difference between the MSSW top list and the severity top list (the mostly lower red line). We also calculate the set difference between the MSSW top list and the frequency top list (the middle yellow line). Note how the red and yellow lines are much closer together than in Figure \ref{fig:MDSE-Set-Difference} and how the red line does not hover close to 0 like it does in Figure \ref{fig:MDSE-Set-Difference}. This demonstrates that the MSSW equation is more evenly balancing inclusion of the top frequency and top severity CWEs.

Note that the goal is not to have the red and yellow lines match. The top list should not necessarily evenly include an equal number of both top frequency and top severity CWEs. Our point with this analysis is to show how the MDSE equation almost exclusively chooses the top frequency CWEs and how our MSSW equation factors in CWEs from both sets. The next subsection will evaluate this more equal inclusion in more detail, focusing on top lists of size 20.

\subsubsection{Chosen CWE Experiment} 

\begin{figure}
\centerline{\includegraphics[scale=.6]{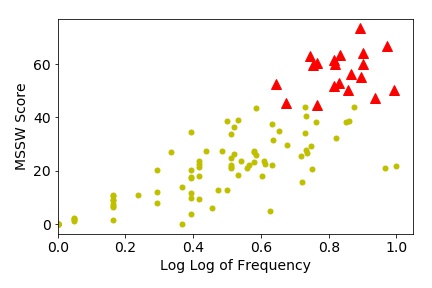}}
\caption{CWEs Chosen (Red Triangles) and Not Chosen (Yellow Circles) for a MSSW Top 20 List Relative to Frequency}
\label{fig:MSSW-vs-Frequency}
\end{figure}

\begin{figure}
\centerline{\includegraphics[scale=.6]{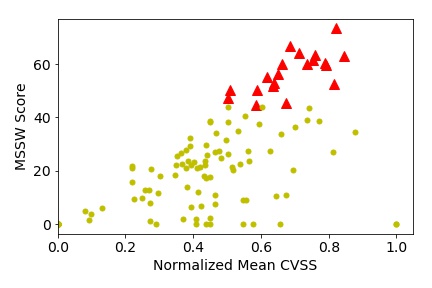}}
\caption{CWEs Chosen (Red Triangles) and Not Chosen (Yellow Circles) for a MSSW Top 20 List Relative to Severity}
\label{fig:MSSW-vs-CVSS}
\end{figure}

Figure \ref{fig:MSSW-vs-Frequency} shows the MSSW scores plotted against the double log frequency $F''_i$ scores. Each point represents a CWE. The red triangles indicate the CWEs that were chosen for the MSSW top 20 list. Note how unlike in the analogous Figure \ref{fig:MDSE-vs-Frequency} for MDSE, there are many higher frequency CWEs that are not chosen for the top 20 list due to their severity not being high enough.

Likewise, Figure \ref{fig:MSSW-vs-CVSS} shows the MSSW scores plotted against the $S'_i$ normalized mean CVSS score for each CWE. Note how the range spreads from 0 to 1, unlike the analogous Figure \ref{fig:MDSE-vs-CVSS} for the MDSE equation. Also note how the MSSW equation chooses CWEs for the top 20 list from CWEs with generally higher CVSS scores. However, it excludes many high severity CWEs because their frequencies were too low. 


\section{2019 Top 20 Lists of the Most Significant Weaknesses}
\label{sec.TopWeaknesses}

We now use our MSSW equation to generate lists of the most significant software security weaknesses. We choose a list size of 20, somewhat arbitrarily, to enable the lists to conveniently fit on a page. We performed the experiments on a variety of list sizes and did not discover any appreciable differences. We did not choose a size of 25 to match the CWE top list because we needed to produce two top lists of differing levels of abstraction to get the most accurate results (explained below) and thus were unable to produce a single list of 25 CWEs for an ideal comparison with the CWE top 25 list.

We follow the approach in \cite{mell2020} of separately providing a top list for CWEs of higher levels of abstraction (pillars and classes) apart from a list covering CWEs of lower levels of abstraction (bases, variants, and compounds). This was done in \cite{mell2020} to avoid errors in frequency calculations that exist in CWE's top 25 list. The paper argues that CWEs mapped to lower level abstractions (e.g., bases) also should count towards their parent abstractions (e.g., classes); this is not done with MDSE calculations. For example, class CWE-20 (Improper Input Validation) is a parent of base CWE-1289 (Improper Validation of Unsafe Equivalence in Input). If a vulnerability exists with CWE-1289, then CWE-20 also needs to be taken into account with the CWE frequency counts. However when this frequency propagation is performed, combining together the two abstractions results in a single top list with a bias towards parents with many children (especially popular children). Thus the 2 levels of abstraction need to be presented in separate top lists.

We will refer to the higher level abstraction list as the class list and the lower level abstraction list as the base list for convenience and because both lists are primarily composed of either classes or bases. We also follow \cite{mell2020} in using published CWE taxonomy views 1000 and 1008 (discussed in Section \ref{CWE}) to propagate CVE data from child CWEs to their parents (discussed above). This provides a more accurate mapping of CVEs onto the CWEs, providing a more accurate data foundation upon which to apply our MSSW equation.\footnote{This propagation especially helps the formulation of the class list since most classes have children. It has a lesser effect on the base list. Note that it is impossible to inverse the propagation of data. CWEs are labelled as specifically as possible by NVD analysts so CVEs described by pillar or class CWEs do not get reflected in the base list. It is even possible that they shouldn't because there may be unidentified bases missing from view 1003 that are still covered by the view 1003 classes.}

These modifications also alter the frequency and severity distributions which could potentially render our double log function invalid. However, Table \ref{tab:correlation} shows correlation results for using and not using all combinations of the modifications adopted from \cite{mell2020}. It shows that the MDSE equation is highly correlated to frequency (.97 or higher) with very little correlation to severity (.25 or lower) regardless of the modifications used or not used. It also shows that the MSSW equation is strongly correlated to both frequency (.81 or higher with one exception) and severity (.66 or higher) regardless of the modifications used. Our one exception is for the class list using propagation with MSSW; even here the frequency correlation was .55 (still strong but much less than the others).

Note that our objective is not for the correlations to necessarily be equal, but that there exists a strong correlation for both frequency and severity. Depending upon the data, the higher frequency CVEs may or may not also be the highest severity CVEs. If so, then the correlations to frequency and severity would both be very high and almost equal. If not, both should still be high but one may be higher than the other. What we do not want in these results is for one of frequency or severity to have a high correlation and the other to have a very low correlation (which can be seen with the MDSE equation).

\begin{table}
\centering
\caption{Measurements Showing the Pearson Correlation of MDSE and MSSW to Frequency and Severity}
\label{tab:correlation}
\begin{tabular}{lllll}
 &  &  & \multicolumn{2}{c}{Correlation} \\
Equation & Abstraction & Propagation & Frequency & Severity \\ \hline
MDSE & All & Yes & .99 & .08 \\
MDSE & All & No & .98 & .18 \\
MDSE & High & Yes & .99 & .10 \\
MDSE & High & No & .98 & .25 \\
MDSE & Low & Yes & .97 & .20 \\
MDSE & Low & No & .97 & .18 \\ \hline
MSSW & All & Yes & .81 & .70 \\
MSSW & All & No & .86 & .66 \\
MSSW & High & Yes & .55 & .96 \\
MSSW & High & No & .84 & .68 \\
MSSW & Low & Yes & .84 & .67 \\
MSSW & Low & No & .83 & .68
\end{tabular}
\end{table}

\begin{figure}
\centerline{\includegraphics[scale=.6]{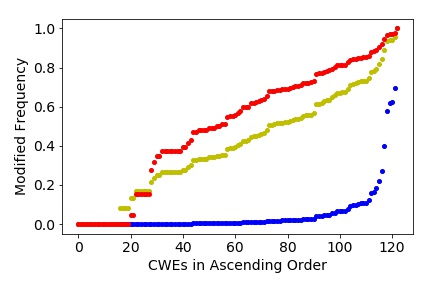}}
\caption{Normalized Distributions of Frequency (bottom blue line), Log of Frequency (middle yellow line), and Double Log of Frequency (top red line).}
\label{fig:G2-distributions}
\end{figure}

We also checked to see that the double log still linearized the frequency distribution when using both variants from \cite{mell2020}. While propagating CVEs over the CWEs using the CWE taxonomies and using all applicable CWEs (i.e., pillars, classes, bases, variants, and compounds), the results show that the double log does still linearize the frequency (see Figure \ref{fig:G2-distributions}). The same results were obtained while also performing the experiment using just the pillars/classes and then just the bases, variants, and compounds (graphs not shown).

Using our MSSW equation to aggregate the frequency and severity of CWEs, the top 20 class list for 2019 is shown in Table \ref{tab:Top20Classes}. The top 20 base list is shown in Table \ref{tab:Top20Bases}. These two lists use the modification from \cite{mell2020} where the CVEs are propagated up through the CWE taxonomies.
We claim that these two lists represent the most accurate measurement yet produced for determining the most significant software security weaknesses. Given that there is no ground truth for how to best combine frequency and severity and no ground truth upon which to establish the correctness of the CVSS metric, it is likely impossible to prove any such metric as maximally effective. We make our `most accurate measurement yet' claim based on the demonstrated limitations in the published MDSE equation and a lack of competing published alternatives.

\begin{table*}
\centering
\caption{2019 MSSW Top 20 Pillars/Classes, Propagating CVSS Data over CWE Taxonomies}
\begin{tabular}{llllll}
Rank & Identifier & CWE Description & MSSW Score & Frequency & Mean CVSS\\
1 & \cwelink{913} & Improper Control of Dynamically-Managed Code Resources & 78.31 & 188 & 8.81 \\
2 & \cwelink{119} & Improper Restriction of Operations within Bounds of a Memory Buffer & 71.14 & 2745 & 8.00 \\
3 & \cwelink{669} & Incorrect Resource Transfer Between Spheres & 64.86 & 181 & 8.31 \\
4 & \cwelink{672} & Operation on a Resource after Expiration or Release & 64.56 & 876 & 7.96 \\
5 & \cwelink{330} & Use of Insufficiently Random Values & 63.74 & 111 & 8.43 \\
6 & \cwelink{704} & Incorrect Type Conversion or Cast & 62.55 & 54 & 8.68 \\
7 & \cwelink{287} & Improper Authentication & 59.75 & 627 & 7.86 \\
8 & \cwelink{345} & Insufficient Verification of Data Authenticity & 54.60 & 483 & 7.73 \\
9 & \cwelink{682} & Incorrect Calculation & 51.94 & 215 & 7.78 \\
10 & \cwelink{269} & Improper Privilege Management & 50.57 & 258 & 7.70 \\
11 & \cwelink{610} & Externally Controlled Reference to a Resource in Another Sphere & 48.38 & 725 & 7.46 \\
12 & \cwelink{706} & Use of Incorrectly-Resolved Name or Reference & 39.04 & 358 & 7.23 \\
13 & \cwelink{20} & Improper Input Validation & 38.56 & 3960 & 6.99 \\
14 & \cwelink{116} & Improper Encoding or Escaping of Output & 32.13 & 2461 & 6.82 \\
15 & \cwelink{400} & Uncontrolled Resource Consumption & 32.07 & 272 & 7.01 \\
16 & \cwelink{74} & Improper Neutralization of Special Elements in Output ... ('Injection')  & 32.06 & 2455 & 6.82 \\
17 & \cwelink{754} & Improper Check for Unusual or Exceptional Conditions & 32.05 & 264 & 7.01 \\
18 & \cwelink{326} & Inadequate Encryption Strength & 28.21 & 35 & 7.24 \\
19 & \cwelink{668} & Exposure of Resource to Wrong Sphere & 26.59 & 2292 & 6.66 \\
20 & \cwelink{436} & Interpretation Conflict & 22.40 & 17 & 7.19 \\
\end{tabular}
\label{tab:Top20Classes}
\end{table*}

\begin{table*}
\centering
\caption{2019 MSSW Top 20 Bases/Variants/Compounds, Propagating CVSS Data over CWE Taxonomies}

\begin{tabular}{llllll}
Rank & Identifier & CWE Description & MSSW Score & Frequency & Mean CVSS\\
1 & \cwelink{89} & Improper Neutralization of Special Elements used  ...  ('SQL Injection') & 71.70 & 384 & 8.89 \\
2 & \cwelink{502} & Deserialization of Untrusted Data & 61.73 & 83 & 9.01 \\
3 & \cwelink{787} & Out-of-bounds Write & 61.57 & 423 & 8.34 \\
4 & \cwelink{78} & Improper Neutralization of Special ... ('OS Command Injection')  & 61.22 & 194 & 8.58 \\
5 & \cwelink{120} & Buffer Copy without Checking Size of ... ('Classic Buffer Overflow') & 59.35 & 162 & 8.55 \\
6 & \cwelink{94} & Improper Control of Generation of Code ('Code Injection') & 58.62 & 100 & 8.72 \\
7 & \cwelink{798} & Use of Hard-coded Credentials & 58.07 & 89 & 8.75 \\
8 & \cwelink{434} & Unrestricted Upload of File with Dangerous Type & 57.95 & 167 & 8.46 \\
9 & \cwelink{416} & Use After Free & 56.69 & 426 & 8.09 \\
10 & \cwelink{352} & Cross-Site Request Forgery  (CSRF) & 51.60 & 386 & 7.86 \\
11 & \cwelink{346} & Origin Validation Error & 51.51 & 430 & 7.82 \\
12 & \cwelink{613} & Insufficient Session Expiration & 51.08 & 402 & 7.82 \\
13 & \cwelink{190} & Integer Overflow or Wraparound & 48.79 & 164 & 7.95 \\
14 & \cwelink{415} & Double Free & 43.17 & 46 & 8.15 \\
15 & \cwelink{125} & Out-of-bounds Read & 42.34 & 658 & 7.28 \\
16 & \cwelink{129} & Improper Validation of Array Index & 41.97 & 25 & 8.50 \\
17 & \cwelink{611} & Improper Restriction of XML External Entity Reference & 41.47 & 100 & 7.69 \\
18 & \cwelink{918} & Server-Side Request Forgery (SSRF) & 41.05 & 74 & 7.78 \\
19 & \cwelink{22} & Improper Limitation of a Pathname to a Restricted ... ('Path Traversal') & 39.40 & 309 & 7.27 \\
20 & \cwelink{191} & Integer Underflow (Wrap or Wraparound) & 37.76 & 18 & 8.47 \\
\end{tabular}
\label{tab:Top20Bases}
\end{table*}


\section{Discussion and Analysis of the Most Significant Weaknesses}
\label{sec.Evaluation}

In this section, we evaluate our 2019 MSSW class and base lists (see Tables \ref{tab:Top20Classes} \& \ref{tab:Top20Bases}) and compare them against the 2019 CWE Top 25 MDSE List \cite{MitreTop25} (reproduced in Table \ref{tab:MITRETop25CWE}). 

\begin{table*}
\caption{Reproduction of the 2019 CWE Top 25 Most Dangerous Software Errors List\cite{MitreTop25}}
\begin{tabular}{llll}
Rank & Identifier & CWE Description & MDSE Score\\

1  & \cwelink{119} & Improper Restriction of Operations within the Bounds of a Memory Buffer                    & 75.56 \\
2  & \cwelink{79}  & Improper Neutralization of Input During Web Page Generation ('Cross-site Scripting')       & 45.69 \\
3  & \cwelink{20}  & Improper Input Validation                                                                  & 43.61 \\
4  & \cwelink{200} & Information Exposure                                                                       & 32.12 \\
5  & \cwelink{125} & Out-of-bounds Read                                                                         & 26.53 \\
6  & \cwelink{89}  & Improper Neutralization of Special Elements used in an SQL Command ('SQL Injection')       & 24.54 \\
7  & \cwelink{416} & Use After Free                                                                             & 17.94 \\
8  & \cwelink{190} & Integer Overflow or Wraparound                                                             & 17.35 \\
9  & \cwelink{352} & Cross-Site Request Forgery (CSRF)                                                          & 15.54 \\
10 & \cwelink{22}  & Improper Limitation of a Pathname to a Restricted Directory ('Path Traversal')             & 14.1  \\
11 & \cwelink{78}  & Improper Neutralization of Special Elements used in an OS Command ('OS Command Injection') & 11.47 \\
12 & \cwelink{787} & Out-of-bounds Write                                                                        & 11.08 \\
13 & \cwelink{287} & Improper Authentication                                                                    & 10.78 \\
14 & \cwelink{476} & NULL Pointer Dereference                                                                   & 9.74  \\
15 & \cwelink{732} & Incorrect Permission Assignment for Critical Resource                                      & 6.33  \\
16 & \cwelink{434} & Unrestricted Upload of File with Dangerous Type                                            & 5.5   \\
17 & \cwelink{611} & Improper Restriction of XML External Entity Reference                                      & 5.48  \\
18 & \cwelink{94}  & Improper Control of Generation of Code ('Code Injection')                                  & 5.36  \\
19 & \cwelink{798} & Use of Hard-coded Credentials                                                              & 5.12  \\
20 & \cwelink{400} & Uncontrolled Resource Consumption                                                          & 5.04  \\
21 & \cwelink{772} & Missing Release of Resource after Effective Lifetime                                       & 5.04  \\
22 & \cwelink{426} & Untrusted Search Path                                                                      & 4.4   \\
23 & \cwelink{502} & Deserialization of Untrusted Data                                                          & 4.3   \\
24 & \cwelink{269} & Improper Privilege Management                                                              & 4.23  \\
25 & \cwelink{295} & Improper Certificate Validation                                                            & 4.06 
\end{tabular}
\label{tab:MITRETop25CWE}
\end{table*}

As stated previously, we expect the MDSE list to vary from the MSSW class and base lists because:
\begin{enumerate}
    \item the MDSE list is biased towards the frequency of a CWE occurring in CVEs,
    \item we use the taxonomy propagation approach from \cite{mell2020}, and
    \item the class and base lists contain a total of 40 CWEs while the MDSE list contains 25 CWEs.
\end{enumerate}



\subsection{High Level Summaries}
View 1003 contains two pillars (\cwelink{682} and \cwelink{697}) and 36 classes, as well as 81 bases, three variants (\cwelink{415}, \cwelink{416}, and \cwelink{401}), and  two compounds (\cwelink{352} and \cwelink{384}). 

The MDSE Top 25 \cite{MitreTop25} ranks CWE items across all the layers of abstraction from view \cwelink{1003}. The list has seven classes, 16 bases, one variant, and one compound. Interestingly, some of these top CWEs have child-parent relationships among themselves. 

A simple inspection of the list shows how parent CWEs do not receive CVE counts from their children.
For example, the count for the top class \cwelink{119} (rank 1, count 1048) does not include the counts of its children \cwelink{125} (rank 5, count 678) and \cwelink{787} (rank 12, count 473). Analogously, the count for the class \cwelink{287} (rank 13, count 299) does not include the counts of its children base \cwelink{798} (rank 19, count 91) and base \cwelink{295} (rank 25, count 77).



Our MSSW class list is comprised of 19 class CWEs and the pillar \cwelink{682} (rank 9) – see Table \ref{tab:Top20Classes}. Only three CVEs are directly described with the pillar, but it appears in the list because there is a set of severe CVEs described with its children (see subsection \ref{NextMostDangerous}). Our MSSW base list is comprised of 17 bases, the variants \cwelink{416} (rank 7) and \cwelink{415} (rank 14), and the compound \cwelink{352} (rank 10) – see Table \ref{tab:Top20Bases}. Each of the two lists properly compare items of the same kind. Interestingly but not surprisingly, each CWE from the base list is a child of a CWE from class list. However, the ordering of these parent-child pairs are not necessarily preserved between the two lists. 


\subsection{Set Differences}
There are differences in the set of CWEs covered by our top 20 MSSW class and base lists and the MDSE list. The pillars/classes in the MDSE list that do not appear in the class list are: \cwelink{200} and \cwelink{732}. The bases/variants/compounds in the MDSE list that do not appear in the base list are: \cwelink{79}, \cwelink{476}, \cwelink{772}, \cwelink{426}, and \cwelink{295}. The base list contains base \cwelink{120} (a child of \cwelink{119}), which does not appear in the MDSE list. 

Note that the two classes from the MDSE list with children in the same list are also in the class list (emphasizing their importance): class \cwelink{119} with children base \cwelink{125} and base \cwelink{787}, and class \cwelink{287} with children base \cwelink{798} and base \cwelink{295}. 

\subsection{Reordered Rankings}
The relative orderings in the MDSE list often do not match the orderings in the MSSW class and base lists.
There are some notable reorderings. \cwelink{89} (Structured Query Language (SQL) Injection) and \cwelink{502} (Deserialization of Untrusted Data) climb up in the base list due to their highest severities of 8.89 and 9.01. \cwelink{913} (Improper Control of Dynamically-Managed Code Resources) does not even appear in the MDSE Top 25 list, as it has only three direct occurrences in the CVEs. However, it climbs up to first position in the class list due to its highest severity of 8.81 and its 188 propagated occurrences. Its main child contributor is base \cwelink{502} with frequency of 83 and severity of 9.01. \cwelink{119} (Improper Restriction of Operations within the Bounds of a Memory Buffer) in the MDSE list, while widely used with 2745 propagated occurrences in the CVEs, is quite less severe than \cwelink{913} and drops to rank 2 in the MSSW class list.

\subsection{The Two Most Dangerous CWEs: Injection vs. Memory Errors}
\label{2MostDangerous}
The two most distinctive groups of weaknesses both in the MDSE Top 25 list and the two MSSW Top 20 lists are injection and memory errors. However, the use of the MSSW equation and the split into  class and base lists considerably reorders these two groups, as well as brings in new CWEs and drops some CWEs.

\subsubsection{Injection Weaknesses}
Injection is the most dangerous type of weakness, represented by  bases) \cwelink{89} (SQL Injection),  \cwelink{502} (Deserialization of Untrusted Data), \cwelink{78} (OS Command Injection), \cwelink{94} (Code Injection), and \cwelink{611} (Improper Restriction of Extensible Markup Language (XML) External Entity Reference), with ranks 1, 2, 4, 6, and 17 respectively in the base list (see Table \ref{tab:Top20Bases}). The MDSE list also contains these five CWEs, however the rankings of the first three are 6, 16, and 11 due to their lower frequencies of 397, 85, and 217. The MSSW inclusion of their high severity scores of 8.89, 9.01,  and 8.58 moved them several positions up in the base list. Note that \cwelink{502} covers Object Injection. 

Also of importance is that the second ranked in the MDSE list \cwelink{79} (Cross-site Scripting), is not in our MSSW base list. Although it has the highest frequency of 1571, its severity score of 5.83 is relatively low. 

The MSSW class list includes \cwelink{913} (Improper Control of Dynamically-Managed Code Resources), \cwelink{116} (Improper Encoding or Escaping of Output), and \cwelink{74} (Injection), ranked 1, 14, and 16 (see Table \ref{tab:Top20Classes}). The reason for that is  \cwelink{913} is the parent of \cwelink{502}, \cwelink{116} is a typical cause of injection and \cwelink{74} is the parent of \cwelink{78}, \cwelink{89}, and \cwelink{94}. Interestingly, the class \cwelink{74} has rank 16 among classes, while its children bases \cwelink{89}, \cwelink{78}, and \cwelink{94} are ranked 1, 4, 6 among bases. The frequencies of 2455 for \cwelink{74}, 384 for \cwelink{89}, 194 for \cwelink{78}, and 100 for \cwelink{94}, leave 1777 injection CVEs that are described with CWEs that are either very infrequent or not severe. These are bases \cwelink{79} (Cross-site Scripting) with the low severity of 5.83, \cwelink{88} (Argument Injection) with the low frequency of 6, and \cwelink{91} (XML Injection) with the low frequency of 16. Being not too dangerous they bring the class \cwelink{74} down to rank 16. That same base \cwelink{79}, not included in the MSSW base list, is ranked 2nd in the MDSE list due to the frequency biasing.

\subsubsection{Memory Weaknesses}

The most dangerous memory weaknesses are \cwelink{787} (Out-of-bounds Write) and \cwelink{120} (Classic Buffer Overflow) with ranks 3 and 5 – see Table \ref{tab:Top20Bases}. Both of them are included in the base list but not the MDSE list, due to the correction of the frequency bias towards proper inclusion of their severity scores of 8.34 and 8.55. 

The other memory weaknesses in the MSSW class and base lists are as follows:
\begin{itemize}
    \item bases \cwelink{125} and \cwelink{787} are buffer overflow (out of bounds read or write)
    \item variant \cwelink{416} is use after free (use of deallocated memory through a dangling pointer)
    \item variant \cwelink{415} is double free (deallocate of already deallocated memory)
    \item class \cwelink{119} is a general memory corruption weakness, which includes buffer overflow, use after free and double free.
    \item class \cwelink{400} is memory overflow (stack/heap  exhaustion) \cite{BFMem}
\end{itemize}


\subsubsection{Injection/Memory Weakness Comparison}
Compared to MDSE, the MSSW equation brings up several injection weaknesses with much higher severity than that of any memory weaknesses. The related CVE analysis confirms that the injection CVE are easier to exploit and with higher impact. An injection directly leads to arbitrary command, code, or script execution. Once a SQL injection is in place, there is no need of additional sophisticated attack crafting or use of glitches in the system. However, it takes considerable extra effort for an attacker to turn a buffer overflow into an arbitrary code execution. He or she would need to have  exceptional skills, such as to apply spraying memory techniques.
The possible damage from an Object injection or from an SQL injection or from is very high. Object injection could lead to remote code execution. An SQL injection may expose huge amounts of structured data, which is proven to be more valuable than raw data. Well formed structured data is easy to read, sort, search, and make sense of it. Via an SQL injection, an attacker could modify a database – insert, update, delete data, execute admin operations, recover file content, and even issue OS commands \cite{OWASPInjection}.

\subsection{Next Most Dangerous CWEs}
\label{NextMostDangerous}
The next most dangerous groups of weaknesses in the MSSW class and base lists relate to file input and upload, authentication,  randomization, cryptography, arithmetics and conversion, and input validation:
\begin{itemize}
    \item randomization – class \cwelink{330} (Use of Insufficiently Random Values) with rank 5 is the class mostly directly assigned to CVEs.
    \item authentication – base \cwelink{798} (Use of Hard-coded Credentials) has rank 7; it is one of the contributors to the class \cwelink{287} (Improper Authentication) with the same rank 7 in the class list.
    \item file upload – base \cwelink{434} (Unrestricted Upload of File with Dangerous Type) has rank 8. It is the main contributors to class \cwelink{669} with rank 3.
    \item cryptography – base \cwelink{352} (Cross-Site Request Forgery) has rank 10, which relates to bugs in data verification. The class list also has class \cwelink{326} (Inadequate Encryption Strength) with rank 18, which is directly assigned to 35 CVEs with severity 7.24.
    \item arithmetics and conversion – base \cwelink{190} (Integer Overflow or Wraparound) and base \cwelink{191} (Integer Underflow) have ranks 13 and 20. They are the primary contributors to pillar \cwelink{682} (Incorrect Calculation) with rank 9. Others in this group on the top lists are bases \cwelink{131} (Incorrect Calculation of Buffer Size), \cwelink{190} (Integer Overflow or Wraparound), and \cwelink{191} (Integer Underflow – Wrap or Wraparound). 
    \item input validation - base \cwelink{129} (Improper Validation of Array Index) has rank 16.
\end{itemize}

\subsection{Mapping Dependencies}
Both the MDSE and MSSW rankings heavily depend on how NVD assigns CWEs to particular CVEs. The CWE selection is restricted to view \cwelink{1003}. Insufficient information about a CVE or an insufficiently specific CWE may lead to the use of the closest matching CWE class or pillar to describe the CVE. For example, it makes sense for class \cwelink{119} to be used for the memory corruption CVE-2019-7098, as there is not much information (no code and no details) – it could be any memory use error or a double free. However, there does exist enough information about the use after free CVE-2019-15554, but it is still mapped to class \cwelink{119} because there exists no appropriate base CWE. A close base CWE is \cwelink{416} (Use After Free), but it does not really reflect memory safe languages like Rust. It is also possible for a class CWE to be assigned to a CVE even when a specific base CWE is available. For example, the stack buffer overflow write CVE-2019-14363 is assigned class \cwelink{119}, although there is plenty of information to map it more specifically to bases \cwelink{121} and \cwelink{120}.


\section{Related Work}
\label{sec.RelatedWork}

The constant need to improve information security has motivated a widespread interest in metrics (both qualitative \cite{guide_secu} and quantitative \cite{ ou2011quantitative}).
As stated by Lord Kelvin, \emph{you cannot improve if you cannot measure}. 
However, many members of the software security community doubt our ability to quantify security. Bellovin was among the first \cite{Bellovin2006} to argue about the infeasibility of software security metrics. \cite{Cox-limit2008} discusses the limitations of the celebrated ``Risk = Threat × Vulnerability × Consequence'' model that is widely used. In \cite{Verendel2009QuantifiedSI} Verendel presents a critical survey of results and assumptions made in the community to quantify security. 
After reviewing over 100 articles, he concludes that the validity of most methods is still strikingly unclear. Many reasons explain this invalidity: lack of validation, lack of comparison against empirical data, and the fact that many assumptions in formal treatments are not empirically well-supported in operational security.

Although we agree, we posit that acceptable but possibly imperfect metrics must be developed in order to facilitate security decisions and to evaluate changes in security posture. To this end, there have been substantial efforts to produce security metrics; \cite{Verendel2009QuantifiedSI} surveys the literature of security metrics published between 1981 and 2008. More efforts can be found in \cite{Purboyo2011}, \cite{Pendleton2017}, and \cite{MorissonSMS}. Security metrics that produce lists of the top security issues are also very prevalent \cite{SymnatecReport2019}, \cite{MCAfeeReport2019}. Specific to software security, there is the OWASP Top 10 \cite{OWASP} for web applications. Also, the CWE project has the Common Weaknesses Scoring System (CWSS) \cite{CWSS} and the Common Weakness Risk Analysis Framework (CWRAF) \cite{CWRAF}, which are used together to provide the most important weaknesses tailored to a particular organization. 

There is also work to critique and improve the foundational data structures used by the MDSE and MSSW metrics. 
CWEs have been discussed in \cite{Wu2015CWE}.
An entirely new approach to classifying software bugs (weaknesses) is proposed by \cite{bojanova2016} and is currently under development.
The evolution of CWE is documented in \cite{CWE-history} (e.g., the addition of classification trees and content for mobile applications and hardware). A critique of CVSS is available in \cite{Jspring2018}. In \cite{mell2020} a novel CWE data collection method is proposed along with simple atomic software security metrics. Our approach in contrast is an aggregate metric designed to be a direct replacement for the MDSE equation.

Along with much other work, our research should be considered as an important step in the process to improve CWE. We believe that our contribution is major as it points out a serious bias in the CWE MDSE equation that is preventing accurate measurements of the most significant software security weaknesses.


\section{Future Work}
\label{sec.FutureWork}

This goal of this work is to identify and fix the unintended bias in the MDSE equation towards frequency. Thus we design the MSSW equation to, as evenly as possible, factor together frequency and severity. And this is rational as it models typical security risk matrices that equally combine probability and impact (e.g., \cite{engert1999risk}). However, it is possible that intentionally biasing towards either frequency or severity is more useful in this domain. Also, the CVSS severity equation is itself an aggregate of exploitability and impact. Future work should evaluate whether or not any intentional bias should be added between these 3 factors.

Also, future work should evaluate additional metrics that might be useful for determining the most significant CWEs. In particular, it would be useful to identify CWEs whose associated vulnerabilities are frequently used in actual and impactful breaches. We note that the CVSS temporal equations provide some of this, but these results are not commonly calculated and no public repository of this data exists. That said, some data does exist to support such mappings (e.g., \cite{Podjarny}).


\section{Conclusion}
\label{sec.Conclusions}

The field of security metrics is a difficult area of scientific research because there is often no ground truth, unlike disciplines such as physics and chemistry. This may lead one to focus on just taking simple low level measurements that are inherently defensible; that was the approach taken in \cite{mell2020}. However, creating aggregate metrics that compose multiple simple measurements is of practical importance for the field of security. In this work we did just that, aggregating frequency and severity (i.e., exploitability and impact) into a single metric. Our objective is not for the correlations to necessarily be equal, but that there exists a strong correlation for both factors which more evenly balances the inclusion of the top frequency and top severity CWEs. This seemingly simple task proved challenging because of the differing distributions of both simpler metrics. Indeed, the officially published CWE metric neglected this property and did not achieve its stated objective (almost exclusively choosing the most frequent CWEs). With our work, we claim to have addressed the limitations and to have produced the most accurate equation yet for measuring the most significant software security weakness.

\bibliographystyle{ACM-Reference-Format}



\end{document}